\documentclass[pra,aps,twocolumn]{revtex4}
\usepackage{epsfig}
\usepackage{amsmath}
\usepackage[dvipsnames]{color}

\begin{document}

\title{Intensity interferometry for observation of dark objects}

\author{Dmitry V. Strekalov, Baris I. Erkmen\footnote{Presently at Google, Inc.%, 1600 Amphitheatre Parkway, Mountain View, CA 94043
} and Nan Yu}
\affiliation{
Jet Propulsion Laboratory, California Institute of
Technology, 4800 Oak Grove Drive, Pasadena, California 91109-8099.}

\date{\today}

\begin{abstract}
We analyze an intensity interferometry measurement carried out with two point-like detectors facing a distant source (e.g., a star) that may be partially occluded by an absorptive object (e.g., a planet). Such a measurement, based on the perturbation of the observed covariance function due to the object's presence, can provide information of the object complementary to a direct optical intensity measurement. In particular, one can infer the orientation of the object's transient trajectory. We identify the key parameters that impact this perturbation and show that its magnitude is equal to the magnitude of the intensity variation caused by the same object. In astronomy applications, this value may be very small, so a differential measurement may be necessary. Finally, we discuss the signal-to-noise ratio that may be expected in this type of measurement.
\end{abstract}


\maketitle

\section{Introduction}

Intensity interferometry has found application in astronomy, specifically in determining the angular diameter of distant stars by measuring the light intensity correlation on Earth. Following the initial wake of excitement caused by the pioneering works by Hanbury Brown and Twiss~\cite{HBT:IntInterf}, the area has shown only limited progress in the later years, chiefly impeded by the stringent requirements that the intensity correlation measurement technique places on photo detectors and supporting electronics, as well as by computationally demanding image processing techniques involved. However, with recent advancement in these technologies, the correlation imaging now experiences an evident revival \cite{Klein2007:SpaceII,Dravins2008:II,Holmes2010:II,LeBoheca2010:II,Holmes2013:CRBII,Dravins12,Dravins13}.
 
A typical observable in intensity interferometry measurements is the Glauber intensity correlation function \cite{Glauber63} which reflects the fourth-order coherence properties of the fields incident on the photo detectors. From this measurement the properties of the source can be learned. For instance, the correlation function full width at half maximum (FWHM), which is also frequently referred to as the speckle or transverse coherence width, yields the angular size of the light source. Moreover, the intensity distribution across a spatially non-uniform source determines the shape of the intensity correlation function (van Cittert - Zernike theorem) and can be extracted from it \cite{Fienup82,Fienup90,Marchesini07,Murray-Krezan12}. This approach has been suggested e.g. for imaging of solar spots \cite{Fienup78}, tidal and rotational distortions, limb darkening \cite{Dravins12,Dravins13} and other stellar phenomena. Resolution of such intensity-correlation imaging corresponds to that of a conventional telescope whose aperture equals the size of the correlated detectors array, and can span kilometers. Intensity interferometry imaging therefore may be compared to a synthetic aperture telescope, however with one important advantage: while in the latter the required timing accuracy of the combined signals is determined by the electromagnetic wave (in our case, optical) \emph{period}, in the former it is determined by its \emph{coherence time}. In particular, this makes intensity interferometers insensitive to atmospheric distortions. Admittedly, this advantage comes at a cost of both the signal-to-noise ratio \cite{Dravins12,Dravins13} and ambiguity in the image reconstruction \cite{Fienup82,Fienup90}.  

In this work we will focus on a generalization of the intensity interferometry approach, wherein our objective is to characterize small changes to the coherence properties of a source (e.g., a star), due to an absorbing object (e.g., a planet) that may be present along the propagation path from the source to the interferometer's detectors \cite{spie}. The presence of an object will change the measured intensity correlation, and this information can be used to estimate some of its features. The specific focus of this paper is on the case of a planet partially occluding a star. 

This paper is organized as follows. In Section~\ref{sc:PF} we introduce a formal description of a distant source partially occluded by a dark object %, similar to the formulations we have used in our prior analysis~\cite{Strekalov13GI}, 
and state the core assumptions and approximations. We will determine the phase-insensitive autocorrelation function of the field incident on the measurement plane, and identify the impact of the object on this autocorrelation, with a few simplifying approximations. In Section~\ref{sc:img} we discuss the intensity covariance estimate obtained by correlating the photocurrents from the two detectors, as a function of their location on the measurement plane. We show how this covariance is modified by the object, and determine the key parameters that impact this signature. We introduce a differential measurement technique that eliminates the prominent coherence signature of the source alone, and isolates the (weak) portion from the object. We apply the results to two examples in two scenarios: (1) a disc-shaped source and object, and (2) a Gaussian-shaped source and object, both in a typical laboratory imaging scenario (a), and in a typical stellar imaging scenario (b). Next, in Section~\ref{sc:SNR} we analyze the signal-to-noise ratio (SNR) that may be expected in the intensity correlation measurements. Finally, in Section~\ref{sc:conc}, we conclude this paper and discuss the results.

\section{Model and approximations}
\label{sc:PF}

\begin{figure}[htp]
\centering
\includegraphics[width = 3.5in]{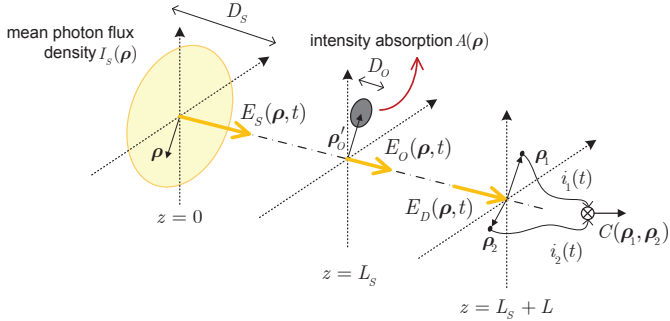}
\caption{Geometry of the problem: $z = 0$ is the source plane, $z = L_s$ is the object plane, and $z = L_s+L$ is the detection plane.}
\label{fig:II}
\end{figure}

Geometry of our model is shown in Fig.~\ref{fig:II}. We use paraxial approximation with the propagation direction denoted as $z$, and assume that the source and the object are two-dimensional. We also assume that the detectors are coplanar. A departure from the latter assumption has been briefly discussed in~\cite{Strekalov13GI} and concluded disadvantageous. A  spatially-incoherent extended source is located at the $z=0$ plane. In this paper we will assume a quasimonochromatic thermal light source with the central wavelength $\lambda$. In practice, this implies that narrow bandpass filters have to be used. We denote the scalar positive-frequency component of the source field as $E_{s}(\vec\rho,t) e^{-i\omega_s t}$, where $\omega_s \equiv 2\pi c/\lambda$ is the center frequency, and $c$ is the speed of light in vacuum. The field amplitude is normalized to the square-root of the photon flux.

For spatially-incoherent thermal radiation, $E_{s}(\vec\rho,t)$ is a zero-mean Gaussian random function that has a nonzero phase-insensitive correlation function~\cite{Mandel,Goodman}
\begin{equation}
\langle E_{s}^{*}(\vec\rho_1,t_1) E_{s}(\vec\rho_2,t_2) \rangle = R(\Delta t) I_s(\vec\rho_1) \lambda^{2} \delta(\vec\rho_2- \vec\rho_1)\,, \label{eq:piscS}
\end{equation}
where $\Delta t = t_2-t_1$, but no phase-sensitive correlation: $\langle E_{s}(\vec\rho_1,t_1) E_{s}(\vec\rho_2,t_2)\rangle = 0$. In Eq.~\eqref{eq:piscS} $\vec\rho_{1,2}$ are two transverse coordinates on the $z=0$ plane, $I_{s}(\vec\rho)$ is the photon flux density in photons$/\text{m}^2/$s, $R(\Delta t)$ is the dimensionless temporal correlation function of the source with $R(0) =1$ and $R(\infty) =0$, and $\delta(\vec\rho)$ is a two-dimensional Dirac delta function. It arises from a delta-function approximation of the spatially-incoherent field's transverse correlation profile, which is appreciable only when $|\vec\rho_2-\vec\rho_1|$ is on the order of a wavelength. We have assumed in Eq.~\eqref{eq:piscS} that the correlation function is separable into the product of the spatial and temporal parts, which is generally true for quasimonochromatic thermal light.

Suppose that a dark object with a finite transverse extent is located at $z =L_{s}$ plane, a distance  $L_{s}$ away from the source. The object modifies the incident field by its transmission function $T(\vec\rho_o)$ which generally may be complex, i.e. may affect both phase and amplitude of the incident light. Then, the field emerging from the object plane is given by
\begin{equation}
E_o(\vec\rho_o,t) = T(\vec\rho_o)\frac{e^{i k  L_{s} }}{i\lambda L_{s}}
\int {\rm d}^2\rho\, E_{s}(\vec{\rho},\tau_s) e^{i k \frac{|\vec\rho_o-\vec{\rho}|^{2}}{2L_{s}} } \label{eq:prxprop1}
\end{equation}
where $\tau_s = t-L_s/c$ and the integration is performed over the \emph{source} plane. Likewise, the field in the detection plane $z=L_{s} + L$ is given by 
\begin{equation}
E_{d}(\vec\rho_m,t) = \frac{e^{i k  L }}{i\lambda L}
\int {\rm d}^2\rho\, E_{o}(\vec{\rho},\tau) e^{i k \frac{|\vec\rho_m-\vec{\rho}|^{2}}{2L} }
\end{equation}
where $\tau = t-L/c$, the integration is performed over the \emph{object} plane and $m=1,2$ represents a detector.

We assume that the detections are performed by two pinhole photo detectors that have equal sensitive areas $A_d$ and quantum efficiencies $\eta$ and are located at $\vec\rho_1$ and $\vec\rho_2$ of the $z = L+L_s$ plane. We also assume that the detectors are small enough to neglect the field variation across $A_d$.  The stochastic photocurrents generated by these detectors as a result of the incident field $E_{d}(\vec\rho,t)$ have the following first-order conditional moments normalized to photoelectrons/s:
\begin{equation}
\langle i_m(t)| E_{d}(\vec\rho_m,t) \rangle = \eta A_d \int {\rm d}\tau |E_d(\vec\rho_m,\tau)|^2 h(t-\tau). \label{eq:immean}
\end{equation}
In Eq.~(\ref{eq:immean}) $h(t)$ is the detectors baseband impulse response, which includes any filtering that follows them prior to the correlation measurement. In order to eliminate a featureless background, it may be convenient to assume that a \textsc{DC} blocking filter is included in $h(t)$, such that $\int {\rm d}t \, h(t) = 0$. 

The blocked \textsc{DC} photocurrent component provides information regarding the total photon flux blocked by the object, which is at the heart of the photon flux based detection methodology, such as used e.g. in the Kepler planetary detection mission~\cite{Fressin11}. Kepler tracks slow intensity variations of a star, to detect Earth-sized exoplanets orbiting the star and to estimate their orbital characteristics. In this work we focus our analysis on the \emph{additional} information that can be gathered via the intensity correlation technique. Note that utilization of this technique does not preclude the observer from also using the mean photon flux registered by each detector.  

The correlation between the intensity fluctuations observed by the two detectors located at $\vec\rho_1$ and $\vec\rho_2$ is estimated by multiplying the two photocurrents and time-averaging the product: 
\begin{equation}
C(\vec\rho_1,\vec\rho_2) \equiv T^{-1} \int_{-T/2}^{T/2} {\rm d}t\, i_1(t) i_2(t), \label{eq:imcorr}
\end{equation}
where $T$ is the multiplication circuit integration time, or the ``coincidence window" if photon counting technique is used. The stationary photo currents correlation measurement converges to a time-independent ensemble average, given by
\begin{equation}
\langle C(\vec\rho_1,\vec\rho_2) \rangle = \mathcal{C} |\langle E_{d}^{*}(\vec\rho_1) E_{d}(\vec\rho_2)\rangle|^{2}, \label{eq:cov}
\end{equation}
where $\mathcal{C} \equiv \eta^2 A_d^2 [|R(t)|^{2} \star h(t) \star h(-t)]$, and $\star$ denotes convolution. For a narrow band source, such that $R(t)$ is much broader than $h(t)$, the parameter $\mathcal{C}$ can be interpreted as a detection volume. For a broadband source this value is reduced proportionally to the square of the $h(t)$ and $R(t)$ widths ratio, that is, to the number $M$ of detected longitudinal modes. This is consistent with a well-known result for Glauber correlation function for a multimode thermal light: $g^{(2)}(0)=1+1/M$. 

Deriving (\ref{eq:cov}) we took advantage of the Gaussian moment factoring of the fourth-order moment of the detected fields~\cite{Mandel}, combined with the assumption that $h_{m}(t)$ blocks \textsc{DC}. Thus, the correlation signature of interest depends on the phase-insensitive correlation function of the detected fields.

We have previously considered the phase-insensitive coherence $\langle E_{d}^{*}(\vec\rho_1) E_{d}(\vec\rho_2)\rangle$ of the detected fields and have been able to write it in an analytical form for a special case of the source luminosity and object absorption both being real Gaussian functions~\cite{Strekalov13GI}. Even though this model was able to roughly approximate the Kepler flux measurement results~\cite{Fressin11}, it is arguably too crude for many objects of interest. Here we will derive a more general expression for the phase-insensitive coherence. 

Immediately after the object the coherence has a form
\begin{equation}
\langle E_o^{*}(\vec\rho_1) E_o(\vec\rho_2)\rangle  = T^*(\vec\rho_1)T(\vec\rho_2)
 e^{i k \frac{\vec\rho_s\cdot\vec\rho_d}{L_{s}}} K_{O}(\vec\rho_d; L_{s}), \label{eq:pisoaprx2} 
\end{equation}
where $\vec\rho_s \equiv (\vec\rho_1+ \vec\rho_2) /2$, $\vec\rho_d \equiv \vec\rho_2-\vec\rho_1$ and
\begin{equation}
K_{O}(\vec\rho; L) \equiv \frac{1}{L^{2}}\int {\rm d}^2\rho^\prime\, I_{s}(\vec{\rho^\prime}) ^{-i k  \vec\rho\cdot \vec{\rho^\prime}/L}.\label{eq:Ko}
\end{equation}

To propagate coherence (\ref{eq:pisoaprx2}) further in the analytical form we need to make approximations. We note that the Fourier transform relation (\ref{eq:Ko}) between $I_{s}$ and $K_{O}$ implies that the latter's width is of the order of $\lambda L_{s}/ D_{s}$, where the source size $D_{s}$ is defined as the diameter over which the photon-flux density is appreciably greater than zero. This width corresponds to a size of the speckle cast by the source onto the object. In many important cases this speckle size is much smaller than the object features we wish to resolve. Then we can write
\begin{equation}
T^*(\vec\rho_1)T(\vec\rho_2)\approx|T(\vec\rho_s-\vec\rho_o)|^2 = 1-A(\vec\rho_s-\vec\rho_o),
\label{eq:T2A} 
\end{equation}
where we have introduced a displacement $\vec\rho_o$ of the object's center from the line of sight and converted the \emph{field} transmission $T$ to \emph{intensity} absorption $A$. Note that in this approximation the phase part of $T$ drops out, so a purely phase object would not alter the coherence propagation within our model.

Approximation (\ref{eq:T2A}) notably simplifies our analysis for propagating the coherence to the detector plane. We derive
\begin{equation}
\langle E_d^{*}(\vec\rho_1) E_d(\vec\rho_2)\rangle =  e^{i k  \frac{\vec\rho_{s}\cdot \vec\rho_d}{L+L_{s}}} K_{O}(\vec\rho_d; L + L_{s}) - K_{D}(\vec\rho_s, \vec\rho_d) \label{eq:soc}
\end{equation}
where the first term is the source's correlation signature in the absence of any object (i.e., free propagation for $L+L_{s}$), and 
\begin{multline}
K_{D}(\vec\rho_s, \vec\rho_d) \equiv \frac{e^{i k  \frac{\vec\rho_s\cdot \vec\rho_d}{L}- i k \frac{\vec\rho_d\cdot\vec\rho_o}{L} }}{\lambda^{2} L^{2}}
\int {\rm d}^2\xi A(\vec{\xi})e^{-i k  \frac{\vec{\rho_d}\cdot \vec{\xi}}{L}} \\
 \int {\rm d}^2\zeta K_{O}(\vec{\zeta}; L_{s})
e^{i k \frac{L+L_s}{LL_{s}} (\vec\rho_o+\vec{\xi})\cdot \vec{\zeta}} 
e^{-i k  \frac{\vec\rho_s\cdot \vec{\zeta}}{L}}  %\label{eq:Kdn} 
\end{multline}
is the modification due to the object. This expression can be simplified using the convolution theorem to 
\begin{multline}
K_{D}(\vec\rho_s, \vec\rho_d) = L^{-2}e^{i k \frac{ (\vec\rho_s-\vec\rho_o)\cdot \vec\rho_d}{L}}\times \\
\int {\rm d}^2\xi I_{s}\left ( \beta (\vec\rho_o+\vec{\xi}) -(\beta-1)\vec\rho_s\right )
A(\vec{\xi})e^{-i k  \frac{\vec{\rho_d}\cdot \vec{\xi}}{L}}, \label{eq:Kdn}
\end{multline}
where $\beta \equiv 1+ L_{s}/L$. 

To continue evaluating the Eq.~\eqref{eq:Kdn} integral we have to make our second important approximation, assuming
\begin{equation}
\frac{D_o}{D_s} \beta \ll 1\,. \label{eq:phsaprx}
\end{equation}
In (\ref{eq:phsaprx}) $D_o$ is the diameter over which the \emph{centered} object's absorption is appreciable. Physically, this means that the angular size of the object (as seen by the observer) is much smaller than the angular size of the source. Let us point out that $\rho_o\beta/D_s\ll 1$ is \emph{not} required, so the approximation~(\ref{eq:phsaprx}) is applicable even to small objects that are far away from the line of sight and to not obscure the source. The signal from such objects is of course vanishingly small. 

When \eqref{eq:phsaprx} holds, we can extract the source intensity from the integral  Eq.~\eqref{eq:Kdn}, arriving at
\begin{align}
K_{D}(\vec\rho_s, \vec\rho_d)  & \approx L^{-2}e^{i k \frac{ (\vec\rho_s-\vec\rho_o)\cdot \vec\rho_d}{L}}\nonumber \\
&\times I_{s}\left ( \beta \vec\rho_o -(\beta-1)\vec\rho_s\right ) )\mathcal{A}\left (\frac{k }{L}\vec\rho_d \right ), \label{eq:socoh2}
\end{align}
where
\begin{equation}
\mathcal{A}(\vec q) \equiv \int {\rm d}^2\rho A(\vec\rho) e^{-i\vec q\cdot \vec\rho}. 
\end{equation}
Substituting Eq.~\eqref{eq:socoh2} into Eq.~\eqref{eq:soc}, and then substituting the result into Eq.~\eqref{eq:cov}, we derive the main analytical result of this work:
\begin{align}
&C(\vec r_s,\vec q_d) \approx \frac{\mathcal{C} }{L^{4} \beta^{4}}\label{eq:covres1} \\
\times &\left | \mathcal{T}_{s} \left (\frac{\vec q_d}{\beta}\right ) - \beta^{2} e^{i \frac{\vec r_s \cdot \vec q_d}{\beta}} 
I_{s}\left (\beta \vec\rho_o-\vec r_s \right ) \mathcal{A}(\vec q_d) e^{-i \vec q_d \cdot \vec\rho_o} \right |^{2}\,, \nonumber
\end{align}
where $\mathcal{T}_{s} (\vec q) \equiv \int {\rm d}^2\rho\, I_{s}(\vec\rho) e^{-i\vec q\cdot \vec\rho}$,  
$\vec q_{d}\equiv k \vec\rho_d/L$, and $\vec r_{s}\equiv L_{s}\vec\rho_s/L$.

\section{Object signature in the correlation function}
\label{sc:img}

Let us consider the case when $\vec\rho_s =0$, i.e. when the two detectors are always symmetrically opposite about the optical axis. In this case, Eq.~\eqref{eq:covres1} simplifies to
\begin{equation}
C(\vec q_d) \approx  \frac{\mathcal{C}}{L^{4} \beta^{4}}  \left | \mathcal{T}_{s} \left (\frac{\vec q_d}{\beta}\right ) - \beta^{2} I_{s}\left (\beta \vec\rho_o \right ) \mathcal{A}\left(\vec q_d\right ) e^{-i\vec q_d \cdot \vec\rho_o} \right |^{2}, \label{eq:covres2}
\end{equation}
where the first term inside the absolute-square is due to the source alone. The second term is the object-induced modification to the correlation function. To quantify the relative magnitude of this object signature we note that $\mathcal{T}_{s}(0)/I_{s}(0)$ and $\mathcal{A}(0)$ are the source and the object effective areas, respectively (or the actual areas, if $I_{s} = const$ for the entire source and $A = 1$ for the entire object). Therefore it is easy to see that the ratio $\beta^2\mathcal{A}(0)I_{s}(0)/\mathcal{T}_{s}(0)\approx(\beta D_o/D_s)^2$ is the fraction of the optical power radiated by the source which is absorbed by the object. This proves an important statement, that \emph{the object signature in the correlation measurement has the same magnitude as in the direct intensity measurement}. Of course, for practical purposes the SNRs in both measurements also need to be compared.

However the main purpose of this work is not to provide a quantitative comparison of the two ways to detect a dark object, but to show that the correlation measurement can yield qualitatively new information, not available from the intensity measurement. At the heart of this capability is the phase between the terms of (\ref{eq:covres2}) which can mediate their constructive or destructive interference. This phase depends on the object displacement $\vec\rho_o$ projected onto the detectors' baseline $\vec\rho_d$, and has no counterpart in the intensity measurements. However leveraging the synthetic aperture analogy discussed in the introduction we notice that this phase variation corresponds to the object passing through Fresnel zones of our fictitious telescope with aperture $\rho_d$. It also should be noted that this exponential arises from a Fourier transform of a shifted object $A(\vec\rho+\vec\rho_o)$ and can be absorbed into $ \mathcal{A}(\vec q_d)$  without loss of generality.

In the following subsections we will consider two analytically tractable examples of objects crossing the line of sight of a thermal light source, in close simulation of the Kepler measurement geometry. We will show that while the intensity variation is obviously independent from the transient direction (e.g., from the planet's orbital plane), the correlation measurement is critically sensitive to this parameter and may serve for its determination.

\subsection{Disc-shaped source and object}

Suppose,
\begin{equation}
I_{s}(\vec\rho) = I_{s}(0)\text{circ}(|\vec\rho|/r_s) \equiv \begin{cases}  I_{s}(0) & \text{ for } |\vec\rho|\leq r_{s} \\ 0 & \text{ otherwise,} \end{cases}
\end{equation}
and 
\begin{equation}
A(\vec\rho) = \begin{cases}  1 & \text{ for } |\vec\rho|\leq r_{o} \\ 0 & \text{ otherwise,} \end{cases}
\end{equation}
where $r_{o} \ll r_{s}$. Substituting these into Eq.~\eqref{eq:covres2}, we can write 
\begin{align}
&C(x,\theta) = \frac{4\mathcal{C}P^2}{L^4 \beta^{4}} \label{eq:covresdisc} \\
\times &\left |\frac{ J_{1}(\pi x/\beta)}{\pi x/\beta} - \beta^2 \gamma^2  \text{circ}(\beta x_o)\frac{J_{1}(\pi \gamma x)}{\pi \gamma x} e^{-i \pi x x_o \cos(\theta)}\right |^{2}\nonumber\,,
\end{align}
where $x \equiv 2 |\vec\rho_d| r_s /(\lambda L)$ is the normalized displacement of the detectors,  $x_o \equiv |\vec\rho_o|/ r_s$ is the fractional displacement of the object relative to the source radius, $\theta \equiv \angle \vec\rho_d - \angle \vec\rho_O$ is the angle between the vectors $\vec\rho_d$ and $\vec\rho_O$, $P \equiv I_{s}(0)\pi r_s^2$ is the mean photon flux of the source, and $\gamma \equiv r_{o}/r_{s}$ is the object-to-source diameter ratio.

Let us consider the image signature from a \emph{differential} measurement between one with no object, and one with the object present, while assuming that nothing else changes. We will also assume that the object is much smaller than the source, $\gamma^2\ll 1$. Then a linearized differential observable is given by the cross-term of Eq.~\eqref{eq:covresdisc} as:
\begin{align}
\Delta C(x,\theta) \approx &-2\mathcal{C}\frac{\gamma}{\beta}\left(\frac{2P}{\pi xL^2}\right)^2\text{circ}(\beta x_o)\label{eq:differential} \\
&\times  J_{1}(\pi x/\beta) J_{1}(\pi \gamma x) \cos\bigl(\pi x x_O \cos(\theta)\bigr).\nonumber
\end{align}

\begin{figure}[htb]
\includegraphics[width=1.7in]{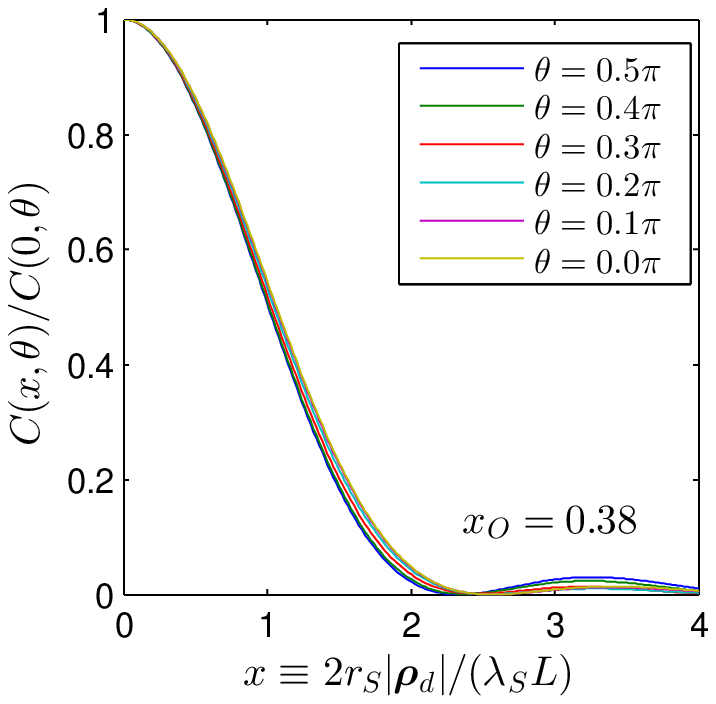}%FMCW_MLEst_MSE_bta0p997_dr0p815_20120817T114906
\includegraphics[width=1.75in]{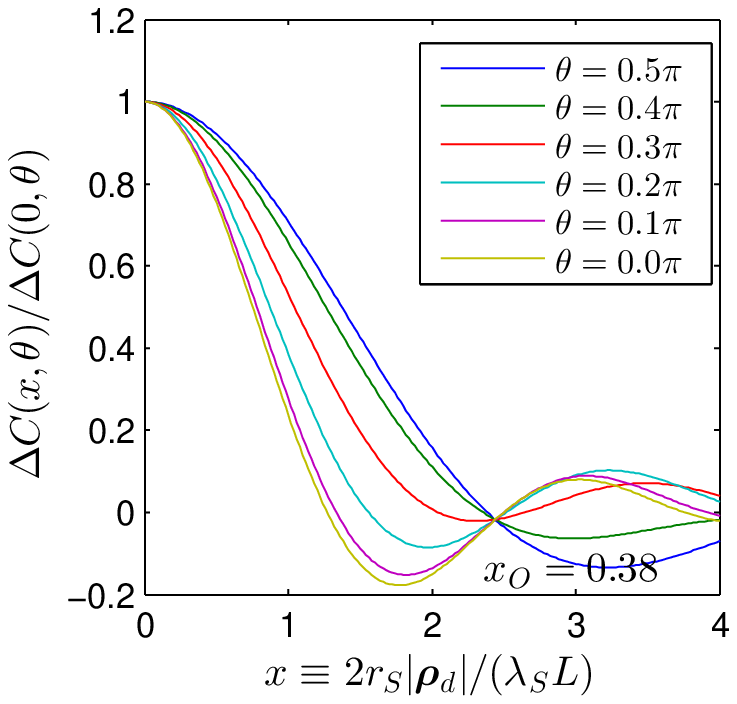}%FMCW_MLEst_MSE_bta0p100_dr0p971_20120820T045257
\\
\includegraphics[width=1.7in]{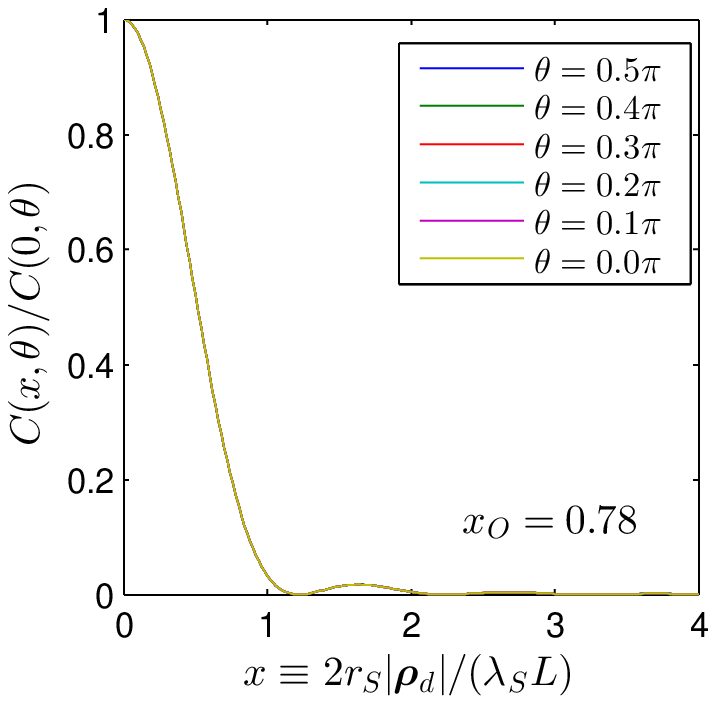}%FMCW_MLEst_MSE_bta0p997_dr0p815_20120817T114906
\includegraphics[width=1.75in]{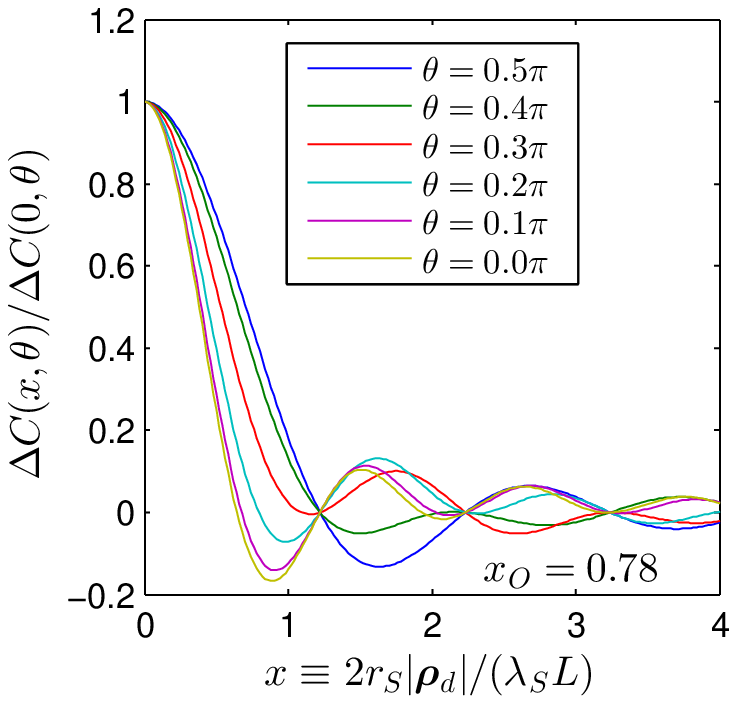}%FMCW_MLEst_MSE_bta0p100_dr0p971_20120820T045257
\caption{The normalized correlation measurement observable $C(x,\theta)$ (left column) and its object-induced variation $\Delta C(x,\theta)$ (right column) for the lab demo case (upper row) and stellar imaging case (lower row) are plotted as a function of $x$ for different $\theta$. The object displacement from the line of sight $x_o$ is fixed as shown.}
\label{fig:discL}
\end{figure}

To evaluate the magnitude of the object's signature we need to specify the parameters of Eq.~(\ref{eq:differential}). Typical values of these parameters are given in Table~\ref{tbl:pars} for two scenarios: a table-top laboratory demonstration, and a Sun-size source partially occluded by an Earth-size planet, being observed from a distance equivalent to that of Kepler 20f. In Fig.~\ref{fig:discL} we show the results for $C(x, \theta)$ and $\Delta C(x,\theta)$ with a fixed object displacement $x_o$, for both lab and stellar cases. 

\begin{table}[t]
\centering
\begin{tabular}{|c|c|c|}
\hline
Variable &  Lab demo & Stellar imaging \\\hline
$\lambda$ [m] & $1\times 10^{-6}$ & $1\times 10^{-6}$ \\
$L_{s}$ [m] & 0.5 & $1.496 \times 10^{11}$ (1 a.u.) \\
$L$ [m] & 0.5 & $8.948 \times 10^{18}$ (290 pc)  \\
$r_{s}$ [m]  & 0.01 & $6.955 \times 10^{8}$ \\
$r_{O}$ [m]  & 0.001 & $6.371 \times 10^{6}$ \\
$\beta$ & 2 & $1+1.67\times 10^{-8}$ \\
$\gamma$ & 0.1 & $9.16\times 10^{-3}$ \\
$\lambda L/(2 r_{s})$ [m] & $2.5\times 10^{-5}$ & $6.433\times 10^{3}$ \\
\hline
\end{tabular} 
\caption{Parameters for a typical lab demo and a stellar imaging example of Kepler 20f.}
\label{tbl:pars}
\vspace*{-0.1in}
\end{table}

\begin{figure}[t]
\includegraphics[width=1.75in]{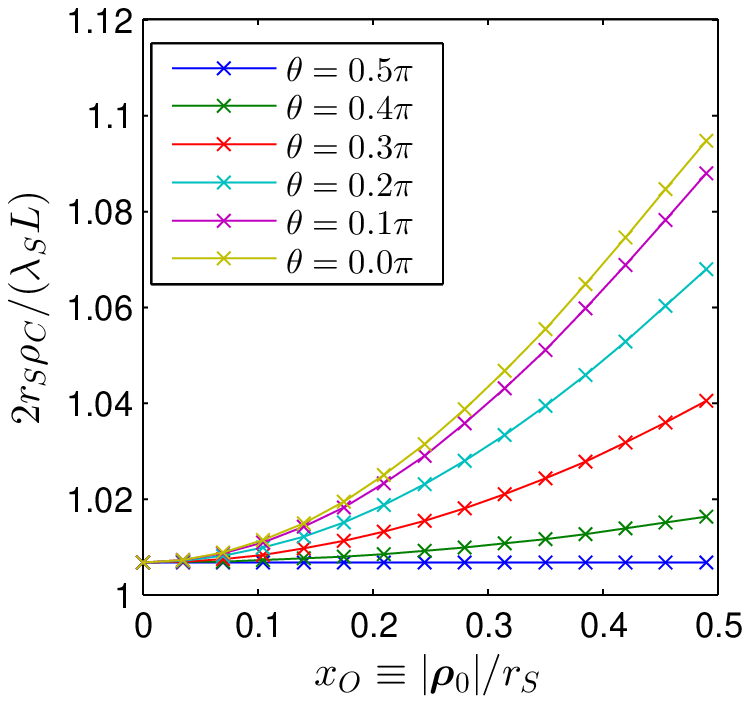}%FMCW_MLEst_MSE_bta0p997_dr0p815_20120817T114906
\includegraphics[width=1.7in]{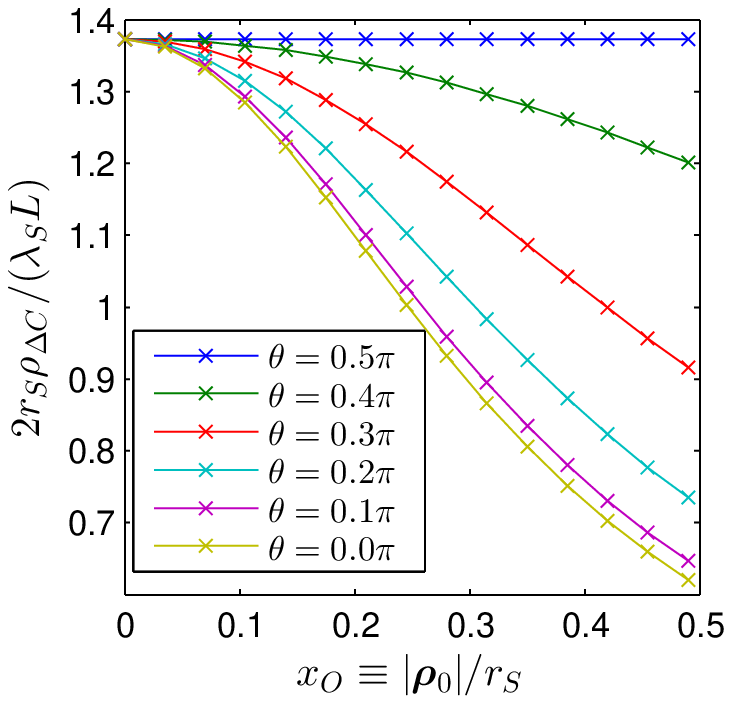}%FMCW_MLEst_MSE_bta0p100_dr0p971_20120820T045257
\\
\includegraphics[width=1.75in]{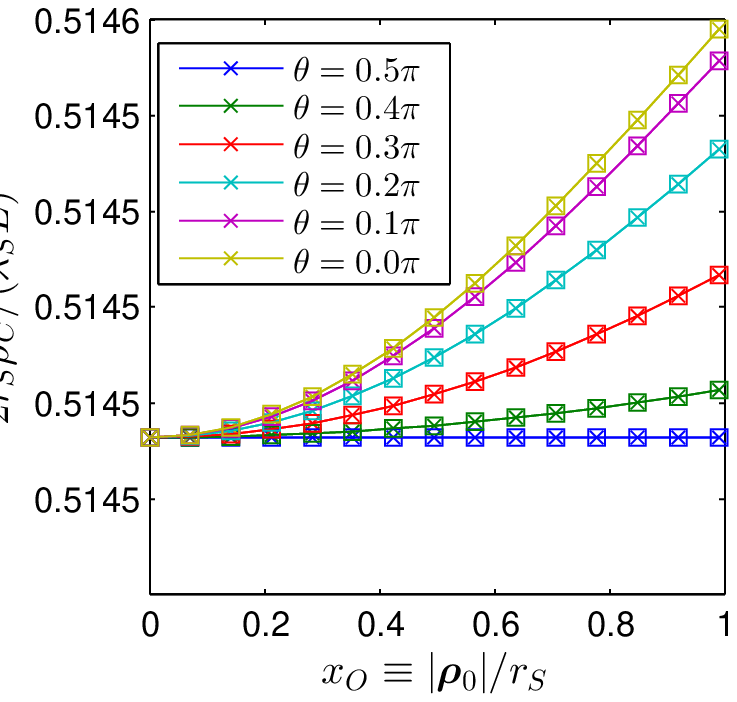}%FMCW_MLEst_MSE_bta0p997_dr0p815_20120817T114906
\includegraphics[width=1.7in]{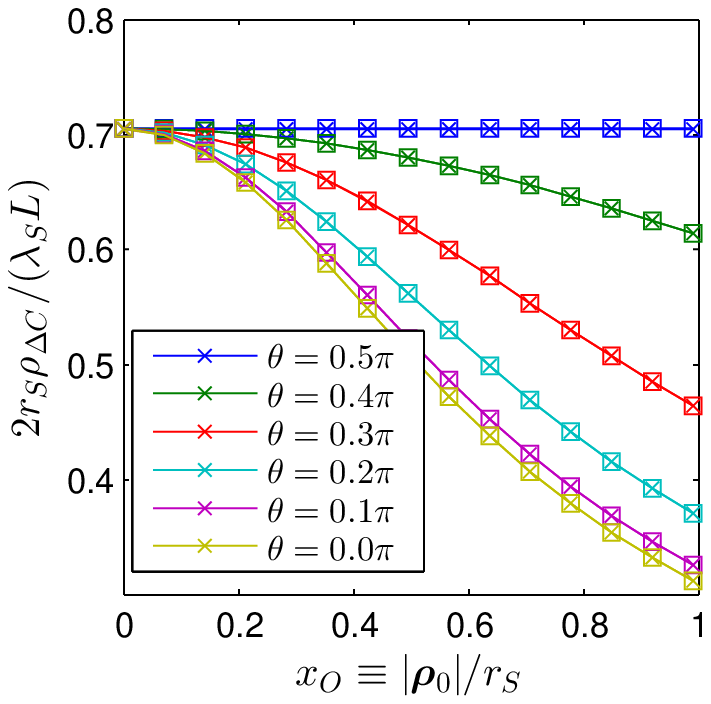}%FMCW_MLEst_MSE_bta0p100_dr0p971_20120820T045257
\caption{Widths of the correlation functions from Fig.~\ref{fig:discL} normalized to the speckle width $\lambda L/(2r_s)$ as a function of the object's transient parameters. Points are obtained from linearized expression~\eqref{eq:differential}, and lines are exact numeric solutions. }
\label{fig:discS}
\end{figure}

From Fig.~\ref{fig:discL} we see that the object signature is mainly manifested by the variation of the correlation function width. We plot this width in Fig.~\ref{fig:discS} as a function of displacement $x_o$ within the range of approximation (\ref{eq:phsaprx}) validity. This plot corresponds to an observation of the object's transient across the source, reaching the line of sight when $x_o=0$. \emph{While the intensity measurement at $x=0$ is obviously independent of this angle, the $\theta$-dependence of the correlation measurement in Fig.~\ref{fig:discS} is evident.} Thus in the stellar imaging example, one would be able to learn about the planetary ecliptic plane orientation from this measurement.

It should be mentioned that while the object shadow observed at any single point does not provide information about the transient direction, the shadow \emph{gradient} may. However it is easy to show (see the Appendix) that in both the Lab demo and especially the Stellar imaging cases the intensity variation across the speckle size due to the shadow is vanishingly small compared to the speckle variation itself. This is because the sharp shadow condition \cite{Laude97} is opposite to assumption \eqref{eq:phsaprx}. Therefore the correlation measurement indeed provides the information unavailable from the intensity measurements.

\subsection{Gaussian-shaped objects}

\begin{figure}[t]
\includegraphics[width=1.7in]{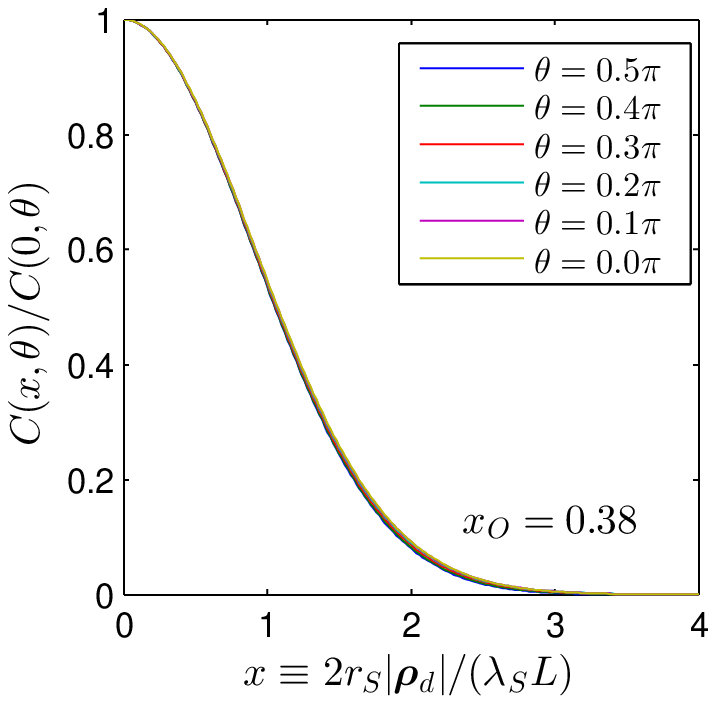}%FMCW_MLEst_MSE_bta0p997_dr0p815_20120817T114906
\includegraphics[width=1.75in]{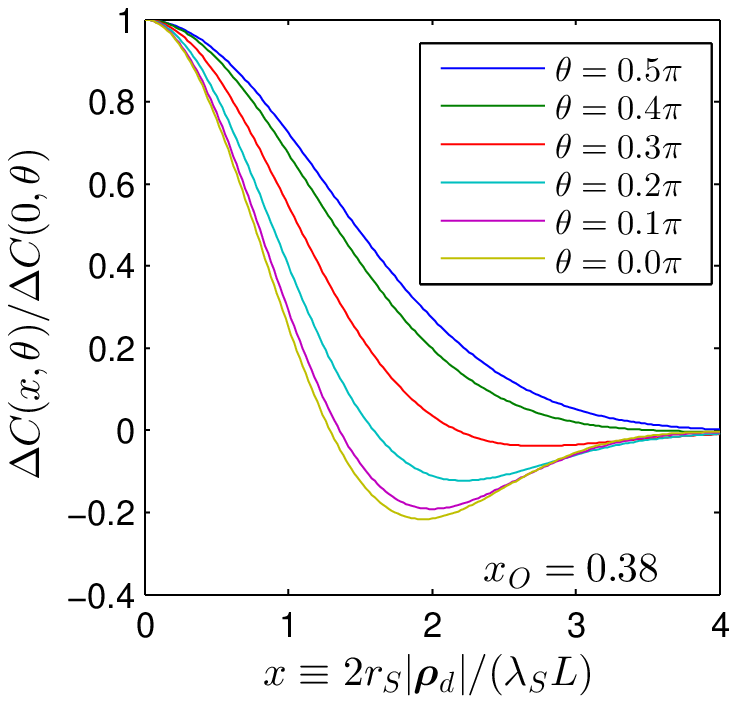}%FMCW_MLEst_MSE_bta0p100_dr0p971_20120820T045257
\\
\includegraphics[width=1.7in]{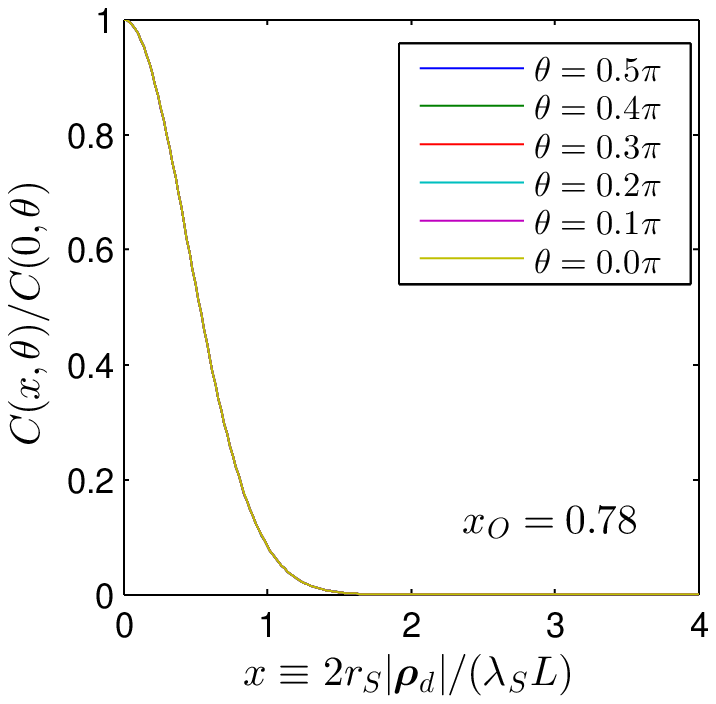}%FMCW_MLEst_MSE_bta0p997_dr0p815_20120817T114906
\includegraphics[width=1.75in]{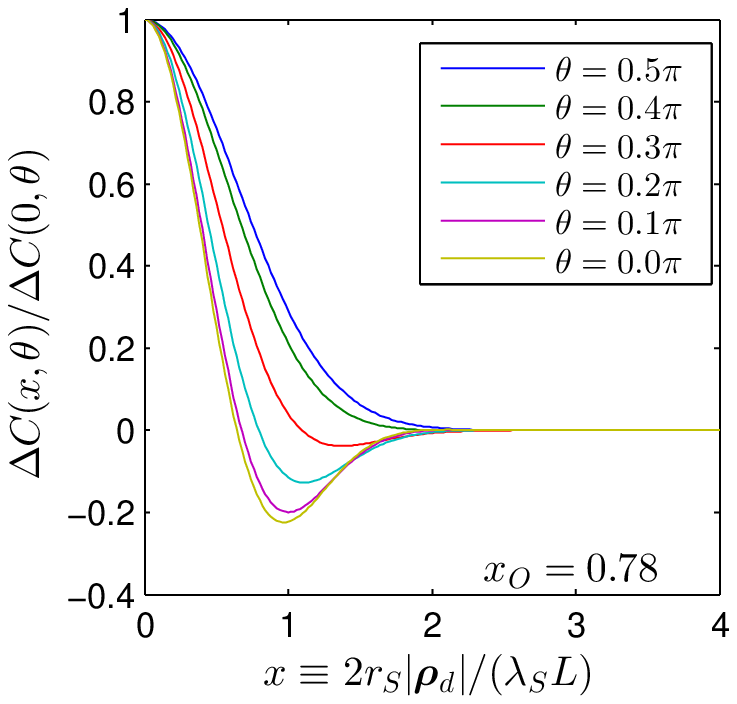}%FMCW_MLEst_MSE_bta0p100_dr0p971_20120820T045257
\caption{Gaussian case equivalent of Fig.~\ref{fig:discL}.}
\label{fig:GausL}
\end{figure}

\begin{figure}[htb]
\includegraphics[width=1.7in]{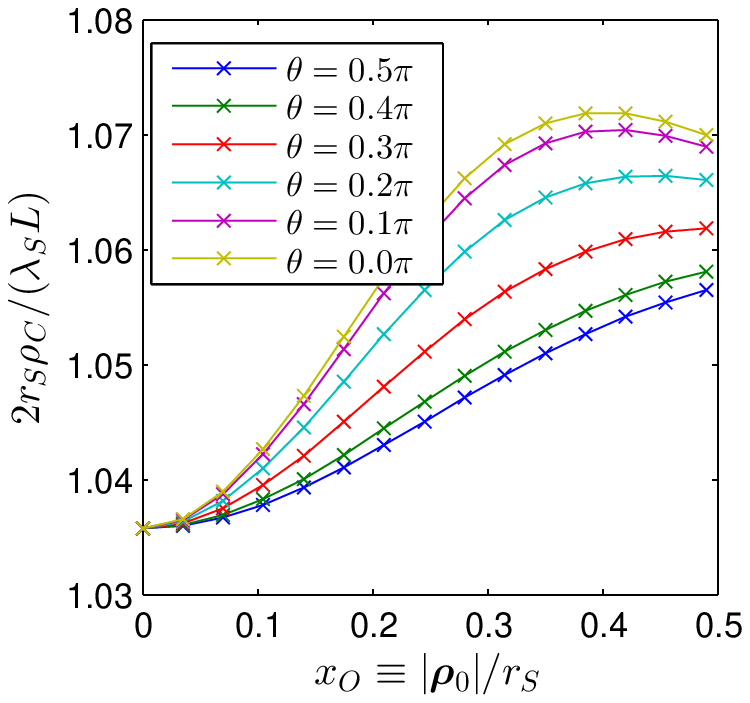}%FMCW_MLEst_MSE_bta0p997_dr0p815_20120817T114906
\includegraphics[width=1.7in]{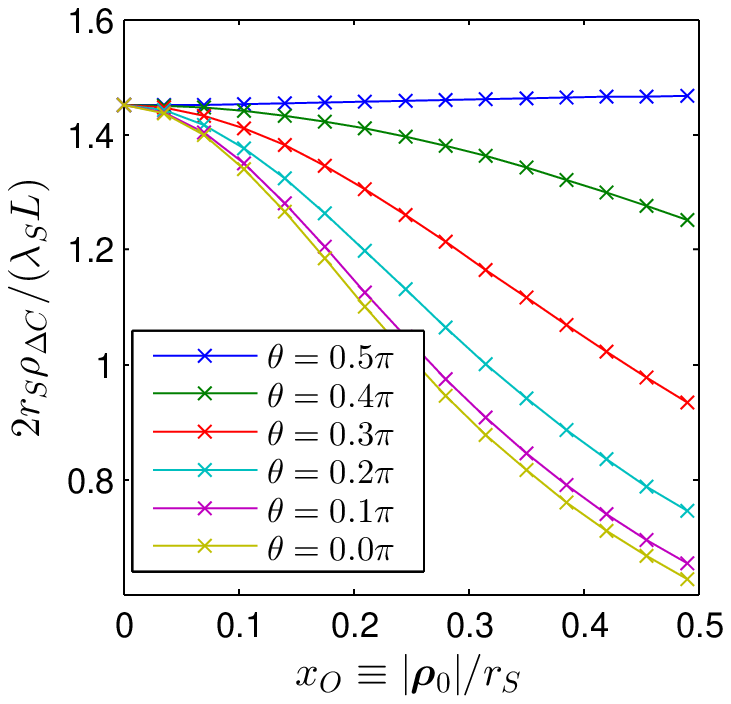}%FMCW_MLEst_MSE_bta0p100_dr0p971_20120820T045257
\\
\includegraphics[width=1.7in]{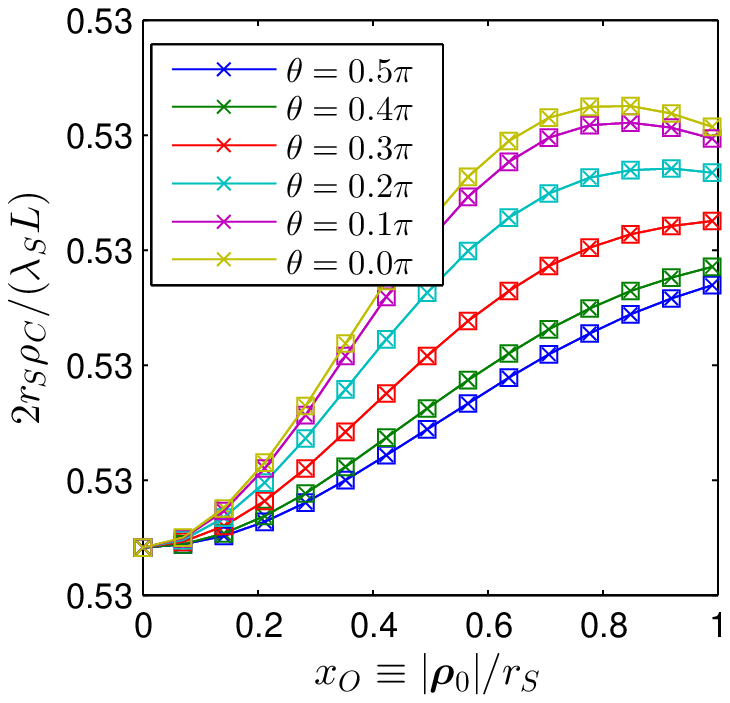}%FMCW_MLEst_MSE_bta0p997_dr0p815_20120817T114906
\includegraphics[width=1.7in]{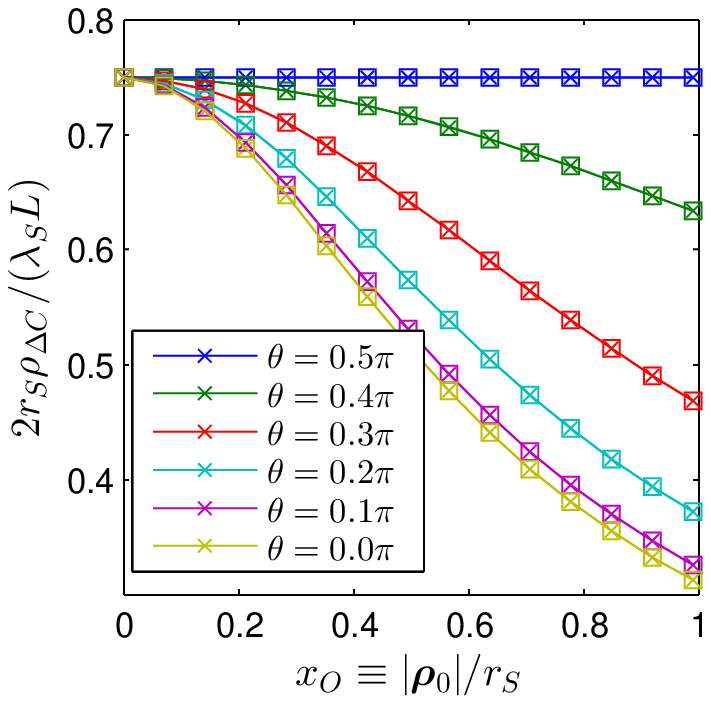}%FMCW_MLEst_MSE_bta0p100_dr0p971_20120820T045257
\caption{Gaussian case equivalent of Fig.~\ref{fig:discS}.}
\label{fig:GausS}
\end{figure}

For the sources and objects that have a Gaussian profile, we have
\begin{equation}
T_{s}(\vec\rho) = e^{-2|\vec\rho|^{2}/r_{s}^{2}}\; {\rm and}\;
A(\vec\rho) = e^{-2|\vec\rho|^{2}/r_{O}^{2}}\,,\label{eq:GSmodel}
\end{equation}
where again $r_{O} \ll r_{s}$. Substituting these into Eq.~\eqref{eq:covres2} and carrying out similar approximations, we obtain
\begin{equation}
C = \frac{\mathcal{C}P_G^2}{L^4 \beta^{4}} \left |e^{-\frac{\pi^2 x^2}{8\beta^2}} - \beta^2 \gamma^2 e^{-2 \beta^2 x_{o}^2} 
e^{-\frac{\pi^2}{8} \gamma^2 x^2} e^{-i \pi x x_o \cos(\theta)}\right |^{2},
\end{equation}
and
\begin{equation}
\Delta C \approx -2 \frac{\mathcal{C}P_G^2}{L^4 \beta^2} \gamma^2 e^{-2 \beta^2 x_{o}^2} e^{-\frac{\pi^2 x^2}{8\beta^2}} 
e^{-\frac{\pi^2}{8} \gamma^2 x^2} \cos\bigl( \pi x x_O \cos(\theta)\bigr)
\end{equation}
where $P_G = I_{s}(0) \pi r_{s}^2/2$, and all other variables have been defined earlier. %Notice an important distinction here is that the contribution of the object-dependent term varies as a continuous function of the displacement $x_o$. %where we have omitted a term proportional to $\gamma^4$.
For the stellar interferometry case with $\beta \approx 1$, we obtain
\begin{equation}
C \approx \frac{\mathcal{C}P_G^2}{L^4 } \left |e^{-\frac{\pi^2}{8} x^2} -  \gamma^2 e^{-2 x_{o}^2} e^{-\frac{\pi^2}{8} \gamma^2 x^2} e^{-i \pi x x_o \cos(\theta)}\right |^{2},
\end{equation}
and
\begin{equation}
\Delta C \approx -2 \frac{\mathcal{C}P_G^2}{L^4} \gamma^2 e^{-2 x_{o}^2} e^{-\frac{\pi^2}{8} (1+\gamma^2) x^2} \cos\bigl(\pi x x_o \cos(\theta)\bigr)
\end{equation}
In Figs.~\ref{fig:GausL} and \ref{fig:GausS} we have plotted the same results as before, but now for the Gaussian case studied here. 


Let us note that despite some quantitative difference between the disk and Gaussian models considered above, they both capture the essential aspects of the object signature. Therefore we can use either the disk model for more realistic approximation of stellar or planetary objects, or Gaussian model for more transparent analytical treatment       .

\section{Signal to noise ratio}
\label{sc:SNR}

In Section~\ref{sc:img} we have shown that, under nominal conditions applicable to a small object obscuring an extended source ($\gamma \ll 1$), the perturbation signature due to the object is weak relative to the baseline signature from the source alone. While a differential measurement can eliminate the source's baseline and improve the visibility of the object's perturbation, it will not eliminate the noise contributed by the source. In this section we derive the signal-to-noise ratio (SNR) of the differential measurement in order to develop a better appreciation for the sensitivity of this measurement and to carry out a valid comparison with the intensity measurement.

Recall  from Section~\ref{sc:PF} that the differential measurement can be expressed as $C_1(\vec{\rho}_1,\vec{\rho}_2) - C_0(\vec{\rho}_1,\vec{\rho}_2)$, where $C_1$ is the Eq.~\eqref{eq:imcorr} measurement \emph{with} the object of interest present, and $C_0$ is the same measurement \emph{without} the object. As typically these two measurements are separated by a duration significantly longer than the coherence time of the photocurrent fluctuations, the two measurements can be assumed statistically uncorrelated. Thus, the variance of the measurement is,  
\begin{equation}
\text{Var}(C_1 - C_0) = \text{Var}(C_1) + \text{Var}(C_0) \approx 2\text{Var}(C_0)
\end{equation}
where the last approximation stems from our earlier observation that the object's perturbation signature is significantly weaker than that of the source when $\gamma \ll 1$. Consequently, in this regime it can be assumed that the variance of either measurement will be dominated by the source-induced shot- and excess-noise fluctuations.

The SNR can, therefore, be expressed as
\begin{equation}
\text{SNR} = \frac{\lvert\langle C_1 - C_0\rangle\rvert^{2}}{\text{Var}(C_1- C_0)} \approx \frac{\lvert\Delta C\rvert^{2}}{2 \text{Var}(C_0)}. \label{eq:snr}
\end{equation}
Let us point out that this definition of SNR apples to the \emph{differential} signal and differs from one in \cite{Dravins12,Dravins13} which would be applicable to measuring $C_0$ or $C_1$ alone.  We have derived the numerator of expression~\eqref{eq:snr} in Section~\ref{sc:img}, thus here we concentrate on the denominator. Using the photocurrent moments discussed above \eqref{eq:immean}, we can express the variance as
\begin{eqnarray}
\text{Var}(C_0) = &&\int {\rm d}\tau_1 \int {\rm d}\tau_2 \int {\rm d}\tau^\prime_1 \int {\rm d}\tau^\prime_2\\
&&K_{h}(\tau_1, \tau_2) K_{h}(\tau^\prime_1, \tau^\prime_2) K_{i}(\tau_1,\tau_2,\tau^\prime_1,\tau^\prime_2)\, ,\nonumber \label{eq:varC0generic}
\end{eqnarray}
where
\begin{equation}
K_{h}(\tau_1,\tau_2) = T^{-1}\int_{-T/2}^{T/2} {\rm d}t\, h(t-\tau_1)h(t-\tau_2)\,, \label{eq:Kh}
\end{equation}
and
\begin{multline}
K_{i}(\tau_1,\tau_2,\tau^\prime_1,\tau^\prime_2) = (\eta A)^2\times \\
\Bigl [ \langle |E_1(\tau_1)|^2 |E_2(\tau_2)|^2  \rangle \delta(\tau_1 - \tau^\prime_1) \delta(\tau_2-\tau^\prime_2)+ \\
+ \eta A \langle |E_1(\tau_1)|^2 |E_1(\tau^\prime_1)|^2 |E_2(\tau_2)|^2 \rangle \delta(\tau_2-\tau^\prime_2) +\\
+ \eta A \langle |E_1(\tau_1)|^2 |E_2(\tau_2)|^2 |E_2(\tau^\prime_2)|^2 \rangle \delta(\tau_1-\tau^\prime_1) +\\
+ (\eta A)^2 \bigl \{ \langle |E_1(\tau_1)|^2 |E_1(\tau^\prime_1)|^2 |E_2(\tau_2)|^2 |E_2(\tau^\prime_2)|^2 \rangle -\\
- \langle|E_1(\tau_1)|^2  |E_2(\tau_2)|^2\rangle \langle |E_1(\tau^\prime_1)|^2  |E_2(\tau^\prime_2)|^2 \rangle \bigr \} \Bigr ].  \label{eq:varEd}
\end{multline}
Here, we have used the short-hand notation $E_m(\tau) \equiv E_{D}(\vec{\rho}_m,\tau)$ for $m=1,2$. The terms in Eq.~\eqref{eq:varEd} have intuitive physical origins: the first term is the  covariance of common-mode fluctuations in the shot-noise (i.e., the conditional variance) from the two detectors, the next two terms are the covariances between the shot-noise fluctuations in one detector and the signal fluctuations in the other detector, and the last term is the covariance between the signal fluctuations (i.e., the conditional mean-square) from the two detectors.

In order to evaluate Eq.~\eqref{eq:varC0generic}, we first perform Gaussian moment factoring~
\cite{Mandel} on each term in Eq.~\eqref{eq:varEd}. This yields expressions for every term in Eq.~\eqref{eq:varEd} in terms of $K_{D}(\vec{\rho}_1,\vec{\rho}_2)$, which is given in Eq.~\eqref{eq:socoh2}. Next, we assume that the \textsc{ac}-coupled photodetector impulse responses $h(t)$ are Gaussian-shaped with $e^{-2}$-bandwidth $\Omega_{B}$, namely,
\begin{equation}
h(t) = \sqrt{\frac{\pi\Omega_{B}^{2}}{2}} e^{-t^2 \Omega_{B}^{2}/8} - \sqrt{\frac{\pi\Omega_{N}^{2}}{2}} e^{-t^2 \Omega_{N}^{2}/8}. 
\end{equation}
The second term here represents the \textsc{dc} notch with bandwidth $\Omega_{N}$. Henceforth, we assume that $\Omega_B \gg \Omega_N$ and $\Omega_N T_0 \ll 1$, which allows to us to effectively neglect the notch's contribution to any nonzero-frequency terms. Our final assumption in evaluating the Eq.~\eqref{eq:varC0generic} is that the integration time $T$ is much longer than the detector's response time ($T\Omega_{B} \gg 1$) and the optical coherence time ($T/T_{0} \gg 1$), such that we may approximate Eq.~\eqref{eq:Kh} as
\begin{equation}
K_h(\tau_1, \tau_2) = T^{-1} \text{rect}\left(\frac{|\tau_1+\tau_2|}{T}\right) [h \star \overleftarrow{h}] (\tau_2 - \tau_1)\,,
\end{equation}
where $\star$ denotes convolution and $\overleftarrow{h}$ denotes time reversal.

Skipping the steps of evaluating each term in the variance expression, we state the final result for the SNR:
\begin{equation}
\text{SNR} = \frac{\cos^{2}(\theta_d)\alpha}{\sigma^2_{ss} + \sigma^{2}_{se} + \sigma^{2}_{ee} }. \label{eq:snr2}
\end{equation}
Assuming symmetric detectors' positions $\vec{\rho}_s=0$, we can write
\begin{equation}
\theta_d = \pi x x_{o} \cos(\theta),
\end{equation}
and
\begin{equation}
\alpha \equiv \biggl \lvert\frac{K_{D}^{(n)}(\vec{\rho}_s,\vec{\rho}_d)}{K_O(0; L+L_S)}\biggr \rvert^{2} = 
\begin{cases} \beta^4 \gamma^4\text{circ}(\beta x_O) \left (\frac{2 J_1(\pi \gamma x)}{\pi \gamma x} \right)^{2}& \\
\beta^4 \gamma^4 e^{-4 \beta^2 x_O^2} e^{-\pi^2 \gamma^2 x^2/4}&  \end{cases} 
\end{equation}
where the upper case correspond to the disk model and the lower case correspond to the Gaussian model discussed above. When $\gamma x \ll 1$, $\beta \approx 1$, and $\beta x_O < 1$ (as in the stellar imaging case), both cases simplify to $\alpha \approx \gamma^4$. 

The three terms in the denominator of the SNR expression are given by
\begin{align}
\sigma^{2}_{ss} &\equiv \frac{\sqrt{2}}{\sqrt{\pi} T \Omega_B \Gamma  N^2 } \left [1+ \frac{ T_0 \Omega_B \Gamma}{\sqrt{8} \sqrt{1 + \frac{\Omega_B^2 T_0^2}{8}}} \right ], \\
\sigma^{2}_{se} &\equiv \frac{2\sqrt{2} }{T  \Omega_B   \Gamma N } \frac{1 + \frac{T_{0}^{2}\Omega_B^{2}}{16}}{\sqrt{1 + \frac{T_{0}^{2}\Omega_B^{2}}{32}}}\times\\
&\left [1+\frac{\sqrt{2}  T_0 \Omega_B \Gamma}{\sqrt{3}}\frac{\sqrt{1 + \frac{T_{0}^{2}\Omega_B^{2}}{32}}}{\sqrt{1 + \frac{T_{0}^{2}\Omega_B^{2}}{8}}\sqrt{1 + \frac{T_{0}^{2}\Omega_B^{2}}{24}}}  \right ],\nonumber \\
\sigma^{2}_{ee} &\equiv \frac{\sqrt{2\pi}}{T \Omega_{B} \Gamma} \sqrt{1 + \frac{T_{0}^{2}\Omega_B^{2}}{16}} \times\\
&\left [1 + \Gamma^2 + \frac{T_0\Omega_B\Gamma}{\sqrt{1+\frac{\Omega_B^2 T_0^2}{8}}}\left(1+\Gamma + \frac{\sqrt{1 + \frac{T_{0}^{2}\Omega_B^{2}}{16}}}{\sqrt{1 + \frac{T_{0}^{2}\Omega_B^{2}}{8}}}\right) \right ]\nonumber.
\end{align}
Here, we have defined $P_S\equiv \int {\rm d}\vec{\rho} I_{s}(\vec{\rho})$ as the mean photon flux of the source, $N \equiv \eta A  T_0 P_{S}/(L+L_s)^2$ as the mean photoelectron count registered per source coherence time, and
\begin{equation}
\Gamma \equiv \left \lvert \frac{K_{O}(\vec{\rho}_d; L+L_s)}{K_{O}(0; L+L_s)}\right \rvert^{2} \in [0,1],
\end{equation}
as the equal-time correlation coefficient between the photocurrents registered at the two detectors, given in terms of $K_O$ defined in Eq.~\eqref{eq:Ko}. 

It is useful to consider two limiting cases of the SNR expression of the incident light being  broadband ($\Omega_B T_0 \ll 1$) or narrowband ($\Omega_B T_0 \gg 1$) relative to the photodetectors bandwidth. Because naturally occurring light sources are nominally broadband and are filtered optically at the measurement plane, typically the former limit will hold. However, with the pseudothermal light sources typically used in the laboratory the latter limit can also be true. 

\begin{figure}[b]
\includegraphics[width=3in]{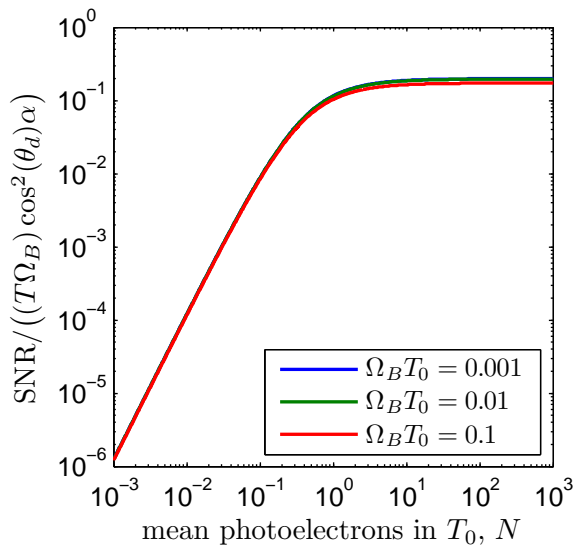}%{SNRbb}
\caption{The normalized signal to noise ratio of the differential intensity covariance measurement is plotted as a function of $N$ for the broadband case. $\Gamma = 1$ is assumed. In this case the normalized SNR has little dependence on the $T_0 \Omega_B$ product. }
\label{fig:SNRBB}
\end{figure}

In the  broadband ($\Omega_B T_0 \ll 1$) limit, the Eq.~\eqref{eq:snr2} expression simplifies to
\begin{equation}
\text{SNR}^{(\text{bb})} \approx \frac{\cos^{2}(\theta_d) \alpha {T \Omega_B \Gamma}}
{\frac{\sqrt{2}}{\sqrt{\pi} N^2} + \frac{2\sqrt{2}}{N} + \sqrt{2 \pi} (1+\Gamma^2) }. \label{eq:SNRbbmax0}
\end{equation}
The photodetector currents decorrelate over approximately $\Omega_B^{-1}$ time interval interval, so the SNR is proportional to $T \Omega_B$. For $N\ll 1$, the signature is photon starved and the SNR has a quadratic dependence on mean photon flux. As $N$ increases, the SNR approaches its maximum value
\begin{equation}
\text{SNR}_{\text{max}}^{(\text{bb})} =    \alpha T \Omega_B \frac{\Gamma\cos^{2}(\theta_d)}{\sqrt{2 \pi} (1+\Gamma^2)}. \label{eq:SNRbbmax}
\end{equation}
Figure~\ref{fig:SNRBB} shows the transition of the normalized SNR from the photon-starved region to its maximum, as a function of $N$.

\begin{figure}[htb]
\includegraphics[width=3in]{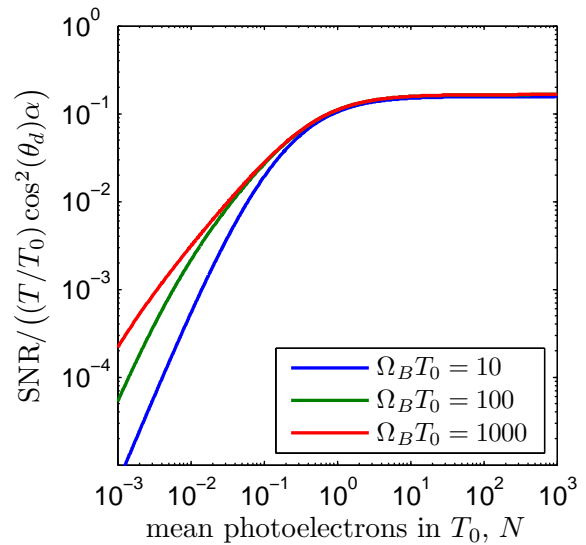}%{SNRnb}
\caption{The normalized signal to noise ratio of the differential intensity covariance measurement is plotted as a function of $N$ for the narrowband case. $\Gamma = 1$ is assumed. In this case the normalized SNR in the $N\ll 1$ regime has a dependence on the $T_0 \Omega_B$ product, but the maximum (attained when $N\gg 1$)is independent of this product.}
\label{fig:SNRNB}
\end{figure}

In the narrowband ($\Omega_B T_0 \gg 1$) limit, on the other hand, the Eq.~\eqref{eq:snr2} expression is
\begin{multline}
\text{SNR}^{(\text{nb})}\approx \cos^{2}(\theta_d) \alpha \Gamma \frac{T }{T_0}
 \bigg[\frac{\sqrt{2}(1+\Gamma)}{\sqrt{\pi} N^2 T_0 \Omega_B}  + \frac{(1+2\Gamma)}{N}\\ +\frac{\sqrt{\pi}}{2\sqrt{2}} \bigl(1+2(\sqrt{2}+1) \Gamma+ (1+2\sqrt{2})\Gamma^2\bigr) \bigg]^{-1}.\label{eq:SNRnb0}
\end{multline}
In this case the photocurrent correlation time is approximately $T_0$, so the SNR is now proportional $T/T_0$. For $N^2 T_0 \Omega_B \ll 1$, the signature is photon starved and the SNR has a quadratic dependence on mean photon flux. As $N$ increases, if $N \sqrt{T_0 \Omega_B} \gg 1$ and $N\ll 1$ simultaneously hold, then the SNR becomes linear in $N$. For $N\gg 1$ it saturates to its maximum value,
\begin{equation}
\text{SNR}_{\text{max}}^{\text{(nb)}} =    \frac{T }{T_0} \frac{2\sqrt{2} \Gamma\cos^{2}(\theta_d) \alpha }{\sqrt{\pi} \bigl (1+2(\sqrt{2}+1) \Gamma + (1+2\sqrt{2})\Gamma^2 \bigr ) } .\label{eq:SNRnbmax}
\end{equation}
Figure~\ref{fig:SNRNB} illustrates the variation of the normalized SNR as a function of $N$ in the narrowband case. 

Let us estimate the SNR for the two examples from Table~\ref{tbl:pars}, assuming $\Gamma = 1$ and $\cos(\theta_d)=1$. In the Lab demo case we assume that a laser-based pseudo-thermal light source is implemented, for which the narrowband limit is appropriate. With such a source $N\gg 1$, can be easily achieved, so we can use the maximum SNR value \eqref{eq:SNRnbmax} which yields $\text{SNR}_{\text{Lab}}\approx 1.65\beta_{\text{Lab}}^4\gamma_{\text{Lab}}^4\,T/T_0=2.64\times 10^{-4}\,T/T_0$. Thus, with a 1 MHz-wide laser we would need on the order of 10 ms integration time to obtain a statistically significant signal. Remarkably, for a better measurement within this scenario one needs a broader band laser (provided that it remains narrowband compared to the detectors).

To evaluate the SNR in the Stellar case, we note that the Kepler 20 is a magnitude 12.497 star in the V+R spectral band \cite{Gautier12}, characterized by the central optical wavelength $\lambda \approx 500$ nm and FWHM spectral range $\Delta\lambda \approx 200$ nm. It produces a photon flux of approximately $5\times 10^5$ photons/s/m$^2$ per normalized spectral interval, given in terms of $\Delta\lambda/\lambda$. Let us assume that we have unity-efficient ($\eta = 1$), fast photo detectors with the bandwidth $\Omega_B = 100$ GHz that are coupled to the same kind of telescopes that were actually used in the Kepler mission, with light collection area of 1.54 m$^2$. In the broad band case we would use the full spectral interval, which gives us the coherence time $T_0=4.17\times 10^{-15}$ s and the photoelectron rate of $3.12\times 10^5$ photons/s. Therefore in the broadband Kepler case $N\approx 1.3\times 10^{-9}$. Substituting this into Eq.~\eqref{eq:SNRbbmax0} and making the same assumptions $\Gamma=1$, $\cos(\theta_d)=1$ as we did for the Lab example, we find
\begin{equation}
\text{SNR}_{\text{Stel}}^{\text{(bb)}} \approx 2\times 10^{-6}\beta_{\text{Stel}}^4\gamma_{\text{Stel}}^4 T,\label{eq:SNRbbstel}
\end{equation}
where the integration time $T$ is in seconds.

Alternatively, we can chose the narrow band measurement strategy and spectrally filter the source radiation so that $\Omega\ll \Omega_B$. It is easy to see that such filtering will not change the spectral brightness of the source, and therefore will not change $N$. Comparing Eqs. \eqref{eq:SNRbbmax0} and \eqref{eq:SNRnb0} in the limit of $N\ll 1$ we then find
\begin{equation}
\text{SNR}_{\text{Stel}}^{\text{(nb)}} = \frac{1}{2}\text{SNR}_{\text{Stel}}^{\text{(bb)}}.\label{eq:SNRnbstel}
\end{equation}
This result indicates that for a uniformly broadband thermal light source the benefit of increasing the correlation function contrast by going into the single-mode detection regime via spectral filtering is negated by the consequent signal reduction. Similar conclusion was reached in \cite{Dravins13} by different reasoning.

\section{Conclusions}         
\label{sc:conc}

We have derived an analytic approximation to the intensity correlation function of an extended source partially occluded by a dark object. We have applied the results to both a table-top demonstration scenario and a stellar imaging scenario. The object signature, defined as a normalized variation of the speckle width, is compared to the direct intensity and intensity gradient signatures in Table~\ref{tbl:results}.

\begin{table}[htb]
\centering
\begin{tabular}{|c|c|c|}
\hline
Observable (normalized)&  Lab demo & Stellar imaging \\\hline
Intensity variation & $9\times 10^{-2}$ & $ 2\times 10^{-4}$ \\
Intensity gradient $\times$ speckle & $ 1.3\times 10^{-4}$ & $ 1.0\times 10^{-17}$  \\
Speckle width variation & $ 7\times 10^{-2}$ & $ 1.7\times 10^{-4}$  \\
\hline
\end{tabular} 
\caption{Magnitude of the object's signature in three types of observables and the parameter sets from Table~\ref{tbl:pars}.}
\label{tbl:results}\vspace*{-0.1in}
\end{table}

We have shown that the magnitudes of the signature expressed in intensity variation and in correlation measurement (the first and the third lines of Table~\ref{tbl:results}, respectively), are very close. The intensity variation however does not reveal any information regarding the direction of the object's transient across the line of sight. Such information could in principle be obtained from the intensity gradient measurement, however the magnitude of this observable (the second line of Table~\ref{tbl:results}) is orders of magnitude smaller. Therefore the intensity correlation measurement provides information unavailable from direct intensity measurements. In particular, a differential measurement (which subtracts the baseline of the unobscured source) yields observable fluctuations that arise from the presence of the object. We have presented our results for two cases of interest, one with disc-shaped objects, and one with objects having a Gaussian profile.

There are several prevailing conclusions to draw from our analysis. First, returning to our key result stated by Eq.~\eqref{eq:covres2},  we point out that the intensity covariance measurement provides information on the magnitude of the Fourier-transform of the absorption profile of the object of interest. Thus, it is possible that with a Gerchberg-Saxton type reconstruction algorithm \cite{Fienup82,Fienup90,Marchesini07,Murray-Krezan12,Fienup78}, one may be able to reconstruct projection shapes of arbitrary dark objects using this signature. In addition, it is clear from our analysis in Section~\ref{sc:img} that the intensity covariance measurement has an imprint of the direction of travel of the object, if several snapshots are taken. Thus, even if full image reconstruction proves too challenging, feature identification of the object seems feasible. 

The major challenge to attaining full images with any algorithmic reconstruction is the signal-to-noise ratio of the measurement. While in a bench-top experiment using a monochromatic pseudo-thermal light source significant SNR can be built up in a very short time, in the stellar case the integration time required to match the SNR of the intensity-based Kepler measurement turns out prohibitively long. This difficulty arizes from the $\gamma^4$ SNR scaling in \eqref{eq:SNRbbstel}, and from a low spectral brightness of natural thermal sources relative to the accessible detectors bandwidths. We should point out that the assumed model of the uniform spectral density may not be correct. Indeed, one might instead expect the presence of bright narrow lines corresponding to atomic transitions. Using such a line as a narrow-band thermal source may greatly improve our SNR expectation \eqref{eq:SNRnbstel}. Further analysis is required to see if this would lead to the SNR values comparable with those in a direct intensity measurement. The main result of this work, however, is the promise of learning \emph{more} about an exoplanet by utilizing measurements that do not require fully resolving the exoplanet with an imaging system.

This work was carried out at the Jet Propulsion Laboratory, California Institute of Technology under a contract with the National Aeronautics and Space Administration. D.V.S. thanks Dr. Igor Kulikov for fruitful discussions.
%%%%%%%%%%%%%%%%%%%%%%%%%%%%%%%%%%% Delete before submitting:
 Copyright 2013 California Institute of Technology. Government sponsorship acknowledged.
%%%%%%%%%%%%%%%%%%%%%%%%%%%%%%%%%

\appendix
\section{Intensity gradient due to a shadow}

To compare our intensity correlation results with direct intensity measurements, let us derive the mean signature obtained by scanning a single pinhole detector at a fixed transverse plane, i.e., let us evaluate
\begin{equation}
\langle i(\vec\rho,t) \rangle  \equiv \eta A \int {\rm d}\tau \langle |E_d(\vec\rho,\tau)|^{2} \rangle h_{\text{lp}}(t-\tau) \label{ddcurr}
\end{equation}
where $E_d(\vec\rho,t)$ is the incident stochastic field at the transverse coordinate $\vec\rho$ and time $t$, $\eta$ is the quantum efficiency of the detector, $A$ is its area, and $h_{\text{lp}}(t)$ is a low-pass filter representing the composite electrical bandwidth of the detector and post-detection processing. The field moment in the integrand is easily obtained by evaluating the right-hand side of Eq.~\eqref{eq:soc}, with the substitutions $\vec\rho_s = \vec\rho$ and $\vec\rho_d = 0$, which yields
\begin{align}
 \langle |E_d(\vec\rho,\tau)|^{2} \rangle &= K_{O}(0; L + L_{s}) - K_{D}(\vec\rho_s, 0)\label{intens} \\
& = \frac{P_{s}}{L^2 \beta^2}\left [ 1 - \beta^2 I_{n}\bigl (\beta \vec\rho_o - (\beta - 1) \vec\rho \bigr)  \mathcal{A}(0)\right ].\nonumber
\end{align}
Here $P_{s} \equiv \int {\rm d}^2\rho I_{s}(\vec\rho)$, and $I_{n}(\vec\rho) \equiv I_{s}(\vec\rho)/I_{s}(0)$ is the normalized source intensity. It is worthwhile to recall that the mean image signature derived herein also requires the assumptions leading to Eqs.~\eqref{eq:pisoaprx2} and \eqref{eq:socoh2} to be valid.

Substituting Eq.~\eqref{intens} into Eq.~\eqref{ddcurr} we assume that $\int {\rm d}t h_{\text{lp}}(t) = 1$ (i.e., unity \textsc{dc} gain) and drop the time variable in the stationary photo current.  We arrive at 
\begin{equation}
\langle i(\vec\rho) \rangle = \frac{\eta A P_{s}}{L^2 \beta^2}\left [ 1 - \beta^2 I_{n}\bigl (\beta \vec\rho_o - (\beta - 1) \vec\rho \bigr) \mathcal{A}(0) \right ]
\end{equation}
as the direct observation signature. Here, the first term is the uniform intensity illumination due to the unobscured source, and the second term is the variation due to the object. The shadow gradient, which may potentially be used for determining the transient direction, can be defined as 
\begin{equation}
\frac{1}{\langle i(\vec{\rho}) \rangle}\frac{\partial\langle i(\vec\rho) \rangle}{\partial\vec{\rho}}
\approx \beta^2(\beta-1) I_{n}^\prime\bigl (\beta \vec\rho_o- (\beta - 1) \vec\rho \bigr) \mathcal{A}(0)\,.\label{eq:grad1}
\end{equation}
For the purpose of the order-of-magnitude estimate, we will assume Gaussian distribution for both the source luminosity and the object opacity (\ref{eq:GSmodel}). Then $\mathcal{A}(0)=\pi r_o^2/2$, and the maximum value of $I_{n}^\prime(\rho_m)=2/r_s$ is achieved at $\rho_m = r_s/2$. 

To make a fair comparison with the intensity interferometry measurement, we need to multiply the gradient (\ref{eq:grad1}) by the measurement baseline, which is of the order of a speckle size $2(L+L_s)/(kr_s)$. We arrive at 
\begin{equation}
\frac{\Delta\langle i\rangle}{\langle i\rangle}\approx\frac{\lambda L_s}{\pi R_s^2}\beta^3\gamma^2.
\label{eq:grad2}
\end{equation}
This expression gives us the values shown in Table~\ref{tbl:results}.

\end{document}